\documentclass{ametsocV6.1}

\usepackage{tikz}
\usetikzlibrary{decorations.pathreplacing}
\usepackage{subfiles}
\usepackage{dsfont}
\usepackage[utf8]{inputenc}
\usepackage{mathtools}
\usepackage{mathrsfs}
\usepackage{hyperref}
\usepackage{subcaption}
\usepackage{pgfplots}
\usepackage{accents}
\usepackage{bbm}
\usepackage{booktabs}
\usepackage{pgfplotstable}
\usepackage{float}

\usepackage{tikz}
\usetikzlibrary{decorations.pathreplacing}

\newcommand{\half}{\frac{1}{2}}

\newcommand{\E}{\mathbb{E}}
\newcommand{\R}{\mathbb{R}}

\newcommand{\1}{\pmb{1}}

\title{Improving Statistical Postprocessing for Extreme Wind Speeds using Tuned Weighted Scoring Rules}

\authors{Simon Hakvoort,\aff{a, b}\correspondingauthor{Simon Hakvoort, s.m.a.hakvoort@gmail.com} 
Bastien Fran\c{c}ois,\aff{a} 
Kirien Whan,\aff{a} 
Sjoerd Dirksen\aff{b} 
}

\affiliation{\aff{a}{Royal Netherlands Meteorological Institute (KNMI), De Bilt, Netherlands}\\
\aff{b}{Mathematical Institute, Utrecht University, Utrecht, Netherlands}\\
}

\abstract{Recent statistical postprocessing methods for wind speed forecasts have incorporated linear models and neural networks to produce more skillful probabilistic forecasts in the low-to-medium wind speed range. At the same time, these methods struggle in the high-to-extreme wind speed range. In this work, we aim to increase the performance in this range by training using a weighted version of the continuous ranked probability score (wCRPS). We develop an approach using shifted Gaussian cdf weight functions, whose parameters are tuned using a multi-objective hyperparameter tuning algorithm that balances performance on low and high wind speed ranges. We explore this approach for both linear models and convolutional neural network models combined with various parametric distributions, namely the truncated normal, log-normal, and generalized extreme value distributions, as well as adaptive mixtures. We apply these methods to forecasts from KNMI's deterministic Harmonie-Arome numerical weather prediction model to obtain probabilistic wind speed forecasts in the Netherlands for 48 hours ahead. For linear models we observe that even with a tuned weight function, training using the wCRPS produces a strong body-tail trade-off, where increased performance on extremes comes at the price of lower performance for the bulk of the distribution. For the best models using convolutional neural networks, we find that using a tuned weight function the performance on extremes can be increased without a significant deterioration in performance on the bulk. The best-performing weight function is shown to be model-specific. Finally, the choice of distribution has no significant impact on the performance of our models.}

\begin{document}

\maketitle

\section{Introduction}

Forecasts of extreme events, such as windstorms, are crucial. These events can have significant implications for society, impacting both industries and individuals who rely on accurate predictions for their planning and decision-making. Given their potential to cause widespread damage \citep[e.g.,][]{Sanou_2023}, improving the prediction of extreme winds are essential.

Weather forecasts are traditionally made using a physics-based numerical weather prediction (NWP) model \citep{Kalnay_2002}, such as the HARMONIE-AROME model \citep{Harmonie} that is run by the Royal Netherlands Meteorological Institute (KNMI). A single run of an NWP model (i.e., a deterministic forecast) contains biases due to errors in the initial conditions and model errors, and is unable to quantify uncertainty in the forecast \citep{PostProcessingOverview}. To address this issue, an ensemble of forecasts can be generated by running the NWP model multiple times with slightly varied initial conditions \citep{EnsembleForecasting, Ensembles_Gneiting}. While this approach can estimate the forecast uncertainty, the forecasts still suffer from systematic biases and errors in the ensemble spread  \citep{EMOS}.

Post-processing involves applying statistical or machine learning methods to the raw outputs of NWP models (either deterministic or ensemble) to correct systematic biases, and generate calibrated probabilistic forecasts. One widely used post-processing approach is \textit{Ensemble Model Output Statistics} (EMOS) \citep{EMOS} that lets parameters of some probability distribution depend linearly on the output of the NWP model. Recent approaches use more advanced non-linear models, such as neural nets \citep{NNEMOS, Daniel, Scheuerer2020, Veldkamp, Hohlein2024, Horat2024}. These models provide more skillful forecasts compared to their linear counterparts. Neural nets, and in particular convolutional neural nets (CNNs) have been successfully been applied to the post-processing of weather data, such as by \citet{Veldkamp} and \cite{Scheuerer2020}. In the former it was used on wind speed data in the Netherlands, and in the latter on precipitation in California, showing performance that is similar to other state of the art methods. However, so far the increase in skill by using neural net-based models occurs mainly for low-to-medium range wind speeds \citep{Veldkamp}. For extreme wind speeds, the methods are typically outperformed by linear EMOS approaches or even climatology. The motivation for our work is to investigate methods to increase the performance of neural net-based EMOS models for extreme wind speeds.

Improving and verifying predictions for extremes is a delicate task. Despite interest from the media and users primarily on extreme events, it is essential that we avoid focusing only on the extremes during both training and verification. This makes sure that we avoid potentially rewarding a biased model (i.e., the \textit{forecaster’s dilemma}, \citet{ForecastersDilemma}) and ensures that our forecasts are not only accurate for extreme events but also reliable for the entirety of the data that is available.

There have been several contributions to improve the EMOS model for extremes. For example, \citet{LogNormalEMOS} and \citet{Lerch_2013_GEV} considered the log-normal (LN) and generalized extreme value (GEV) distributions as forecasting distributions and found that these improved over the truncated-normal (TN) distribution in terms of weighted scoring rules for high thresholds. A truncated GEV can be used to avoid forecasting negative wind speeds \citep{Baran_2021_trunc_gev}. \citet{MixtureEMOS} also suggest the use of mixture and regime switching distributions, for which different distributions can be used depending on a statistic of the ensemble forecast. 

Several authors already proposed to use a loss function that penalizes errors in the extreme range more severely  \citet{hess_deeplearning_adjusted_loss, Scheepens2023}. These studies use neural nets in a deterministic forecasting (i.e., regression) setting. Both find that using a weighted loss function leads to improved performance of the network on extremes, compared to non-weighted counterparts. However, \cite{hess_deeplearning_adjusted_loss} observe that this improvement in performance comes at the expense of increased overcasting and false alarm ratio.

In this work we focus on improving the performance on extremes for EMOS with both linear and convolutional neural net-based models. We focus on wind speed in the Netherlands and use deterministic HARMONIE-AROME forecasts with a lead time of +48 hours as inputs for our postprocessing methods. We investigate two aspects that can improve forecasts for extremes. First, we explore the choice of parametric distribution by modeling wind speed as a TN, LN, GEV, as well as mixtures of these distributions. Building on \citet{MixtureEMOS}, we propose an adaptive mixture distribution, where the weight of the mixture distribution is based on the wind speed forecast of the NWP model. A-priori we expect TN to be better at predicting more moderate wind speeds and LN and GEV to be better for extremes, since these distributions have a heavier right tail compared to the TN and are more suitable for modelling extremes. Second, we use weighted scoring rules for probabilistic forecasts that focus more on particular parts of the distribution. Traditionally, the continuous ranked probability score (CRPS) and the log-score are used as loss functions during training, and we extend this by using the weighted continuous ranked probability score (wCRPS) to train the models \citep{Gneiting_twCRPS_2011}. The wCRPS introduces a weight function into the CRPS to emphasize the importance of certain events. The weight functions considered in this work are vertically shifted Gaussian CDFs, where the parameters in the weight function are tuned using cross-validation. With such a weight function the wCRPS is able to focus more on the accuracy of the forecasts of extremes compared to more moderate events. By training EMOS models using the wCRPS as loss function, we expect to improve the performance on extremes.
 
Independently of our work\footnote{The results of this paper were announced earlier in the first author's thesis \citep{hakvoort2024thesis}.}, \citet{wessel2024improvingprobabilisticforecastsextreme} use a similar strategy. The authors study 10-meter wind speed with an 18-member ensemble forecast in the United Kingdom with a lead time of +48 hours. They train linear EMOS models with both the TN and truncated LN as parametric distributions. They show that using a wCRPS with an indicator function (known as a threshold weighted CRPS (twCRPS)) as the loss function leads to an improvement in performance for extreme wind speeds at the expense of worse performance for moderate wind speeds. The authors call this a `body-tail trade-off'. They propose two approaches to address the issue: the first involves training on a linear combination of the CRPS and twCRPS, while the second approach uses a convex combination of models trained separately on the CRPS and twCRPS. The scope of our research is broader, as we use the wCRPS as a loss function with different weight functions and examine its effect on EMOS with linear models as well as convolutional neural nets. Moreover, we investigate the effectiveness of tuning the weight function in the wCRPS for managing the body-tail trade-off. Additionally, the impact of the choice of loss function on the selection of the parametric distribution is analyzed.
\newline

This article is structured as follows. In section \ref{sec: data} we describe  the used dataset. Next, in section \ref{sec:methods}, we discuss the methods, including a description of the EMOS model and its implementation. We then present the results in section \ref{sec:results} and conclude with a discussion of our findings in section \ref{sec:conclusion and discussion}.

\section{Data}\label{sec: data}

We use deterministic forecasts from the HARMONIE-AROME model (cycle 40, HA40) \citep{Harmonie} between 2015 and 2019. HA40 has a horizontal grid spacing of 2.5 km, and we use forecasts initialized at 0000 UTC with a lead time of +48 hours. We use a set of predictors that were chosen through forward selection in a previous study with the same data set \citep{Daniel}. It is possible that further examination of the predictor set could yield improved results, but this is outside the scope of this study.  
We extract data from all input variables at the grid point closest to the station observations, as described below. Additionally, for wind speeds we also extract a grid with an odd side length around the station, so that the station is located in the center of the box.

\begin{table}[h]
    \centering
    \begin{tabular}{c}
        \toprule
        \textbf{Predictors} \\
        \midrule
        Wind speed at 10 meter height \\
        Mean sea level pressure \\ 
        Turbulent kinetic energy \\
        Humidity at 2 meter height \\
        Geopotential height at 500 hPa \\
        \bottomrule
    \end{tabular}
    \caption{Predictor variables from the HA40 forecast}
    \label{tab:predictors}
\end{table}

\begin{figure}[h]
    \centering
\includegraphics[width=0.7\linewidth]{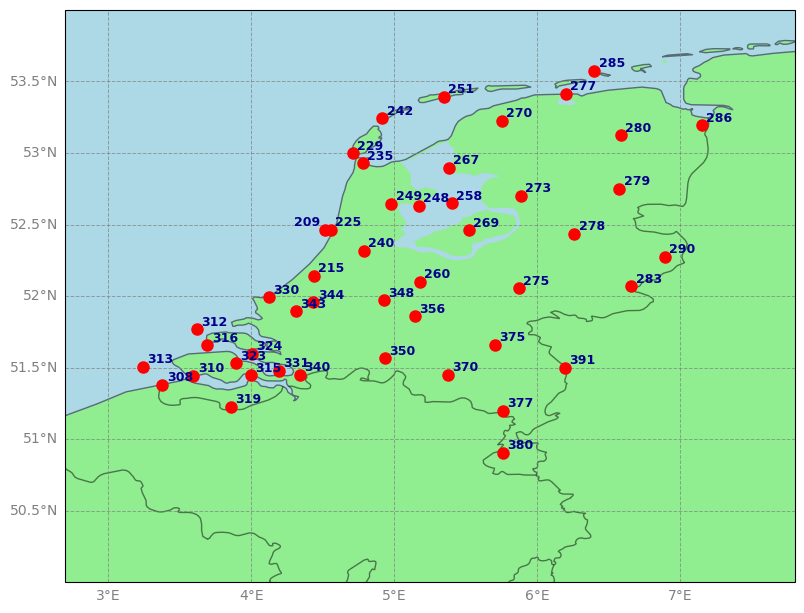}
    \caption{Locations of the Dutch weather stations, including their unique station codes. Table \protect{\ref{tab:station data}} in Appendix B contains more information regarding the stations.}
    \label{fig:locations stations}
\end{figure}

The predictand data is 10-minute average 10-meter winter (November to March) wind speed observations at 47 automatic weather stations maintained by KNMI (Figure \ref{fig:locations stations}). We use all available observations, except for the data from stations 229, 285, and 323 (Texelhors, Huibertgat, and Wilhelminadorp) which are omitted due to a significant amount of missing data.

\begin{figure}[h]
    \centering
\includegraphics[width=0.7\linewidth]{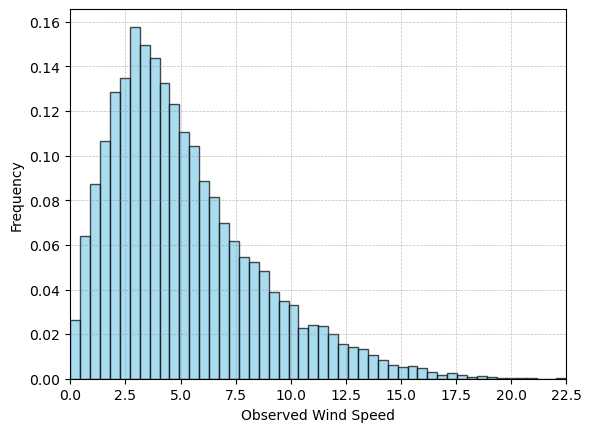}
    \caption{Frequency histogram of the observed wind speeds in the full training data set.}
    \label{fig:histogram wind speeds}
\end{figure}

 The histogram of observed wind speeds in the training set shows the heavy-tailed observations. The bulk of the observations are between 0 and 10 m/s, and there is a long tail for the higher wind speeds (Figure \ref{fig:histogram wind speeds}). 
 
The data is separated by chronological order into four different folds, with three folds used for training and model selection using three-fold cross-validation, and one fold reserved as a separate test set for evaluating model performance (see Table~\ref{tab:data folds}). We choose this splitting strategy to minimize the effect of autocorrelation between different folds. 

\begin{table}[h]
    \centering
    \begin{tabular}{|c||c|}
    \hline
        Fold 1 & October - December 2015 and January - March 2016 \\
        \hline
        Fold 2 & October - December 2016 and January - March 2017 \\ 
        \hline 
        Fold 3 & October - November 2017 and January - March 2015 \\
        \hline
        \hline
        Test Set & November - December 2018, January - March 2019 and October - November 2019 \\
        \hline
    \end{tabular}
    \caption{Folds 1, 2 and 3 are used for model selection and validation. The test set is used to evaluate the performance of the models.}
    \label{tab:data folds}
\end{table} 
Once the final models and hyperparameters are selected using cross-validation, we re-train each model with the selected hyperparameters on all three folds and evaluate the performance of the model on the test set.

Inter-annual variability in the observed wind speed is evident between the folds. We can see that the first fold contains the most extreme observations, while the second fold contains a relatively calm winter period (Table \ref{tab:data quantiles}).  

\begin{table}[h]
    \centering
    \begin{tabular}{lccccc}
    \toprule
       &\# observations & mean & 90$^{th}$ percentile & 95$^{th}$ percentile & 99$^{th}$ percentile \\ 
       \midrule
        Training Data& 23513  & 5.16 & 9.87 & 11.88 & 15.43\\
        \midrule
        Fold 1& 7754 & 5.62  & 10.84 & 12.91 & 16.32 \\
        Fold 2& 7979 & 4.52 &
    8.41 &
10.02 &
13.26  \\
Fold 3 &  7780&5.37 &
10.29 &
12.21 &
15.44 \\
        \bottomrule
    \end{tabular}
    \caption{Wind speed statistics (mean and various percentiles) for the different folds (m/s). All folds are approximately the same size. The row `Training Data' consists of all three folds combined.}
    \label{tab:data quantiles}
\end{table}

\section{Methods}\label{sec:methods}
This section details the models used, focusing on models that predict the parameters of a forecast distribution with linear models and convolutional neural nets. Afterwards, we further discuss the choices of parametric distribution and scoring rules. Then the implementation is described, ending with an explanation of the hyperparameter optimization algorithm.

\subsection{Linear Model}
We will now provide an overview of the linear model (EMOS), which was first proposed by \citet{EMOS}. It is one of the most commonly used statistical post-processing methods, both in academics as well as in operational environments  due to its ease of application \citep{EMOS_at_MET}. We assume that wind speed follows a specific distribution. The distribution parameters are then estimated based on either the deterministic or ensemble mean NWP forecasts. In the standard approach, one uses a linear model for the relation between the predictors $X_1,...,X_n$ and the parameters. Concretely, for a forecast distribution with a location and scale parameter (e.g., the truncated normal distribution), one models

\begin{align}\label{eq:emos mean ensembles}
    \mu = a_0 + \sum^n_{i=1} a_i X_i
\end{align} 
and
\begin{align}\label{eq:emos var features}
    \sigma^2 = \text{softplus}\left(b_0 + \sum^n_{i=1}b_i X_i\right),
\end{align}
where the softplus function, $\text{softplus}(x) = \ln\left(1 + e^x\right)$, is used to enforce a positive scale parameter. The parameters $a, b \in \R^{n+1}$ are chosen by fitting the training data using a specific scoring rule, see below for details. 

During preliminary testing, we also examined different variants to model the scale parameter of the distribution. In particular, we considered using the spatial variance around the station as either the sole predictor or an additional input for the scale parameter. This did not lead to improved performance.

\subsection{Convolutional Neural Networks}

We also estimated the parameters of the forecast distribution using convolutional neural nets (CNNs). CNNs are a type of neural net that works well to handle data with a grid-like structure, such as a grid of pixels for image data. In addition, CNNs can capture complex non-linear relationships between variables, unlike linear models. For more information regarding CNNs, we refer to \citet{Goodfellow}.

\begin{figure}[htbp]
    \centering
    \scalebox{0.5}{\begin{tikzpicture}

\node[anchor=north west, font=\Large] at (-9, 1) {(a)};

\tikzstyle{layer}=[draw, rounded corners, minimum height=1cm, minimum width=6cm, align=center, font=\Large]
\tikzstyle{input}=[layer, fill=blue!20]
\tikzstyle{conv}=[layer, fill=green!20]
\tikzstyle{concat}=[layer, fill=gray!20]
\tikzstyle{dense}=[layer, fill=purple!20]
\tikzstyle{output}=[layer, fill=orange!20]

\node[input] (input1) at (-4, 0) {i)  Wind Speed Forecast ($15\times 15$ grid)};
\node[input] (input2) at (4, 0) {ii) Secondary Inputs (grid point)};
\node[input] (input3) at (8, -10.5) {iii) Wind Speed Forecast (grid point)};

\node[conv] (conv_block) at (-4, -2) {Convolutional Block};

\node[concat] (concat) at (0, -4) {Concatenate};

\node[dense] (dense1) at (0, -6) {Dense, up to 200 Units};

\node at (0, -7) {\vdots};
\node at (0, -7.5) {\vdots};
\node at (0, -8) {\vdots};

\node[dense] (dense2) at (0, -9) {Dense, up to 200 Units};

\node[concat] (concat2) at (0, -10.5) {Concatenate};

\node[output] (output) at (0, -12) {Output};

\draw[thick, ->] (input1.south) -- (conv_block.north);
\draw[thick, ->] (conv_block.south) -- (concat.west);

\draw[thick, ->] (input2.south) -- (concat.east);

\draw[thick, ->] (concat.south) -- (dense1.north);
\draw[thick, ->] (dense1.south) -- (0, -7.25);
\draw[thick, ->] (0, -7.9) -- (dense2.north);
\draw[thick, ->] (dense2.south) -- (concat2.north);

\draw[thick, ->] (input3.west) -- (concat2.east);

\draw[thick, ->] (concat2.south) -- (output.north);

\draw[decorate,decoration={brace,amplitude=10pt,mirror},thick] 
    (3.5,-9.5) -- (3.5,-5.5) 
    node[midway,xshift=2cm,align=left,font=\Large] {up to 5 times};

\begin{scope}[xshift=15cm]

\node[anchor=north west, font=\Large] at (-5, 1) {(b)};

\tikzstyle{input}=[layer, fill=blue!20]
\tikzstyle{conv}=[layer, fill=green!20]
\tikzstyle{bn}=[layer, fill=yellow!20]
\tikzstyle{pool}=[layer, fill=red!20]

\node[input] (input) at (0, 0) {Input Wind Speed Grid};

\node[conv] (conv1) at (0, -1.5) {7x7 Conv, 8 Kernels};
\node[bn] (bn1) at (0, -3) {Batch Normalization};

\node[pool] (pool1) at (0, -5.5) {2x2 Max Pooling};

\node[conv] (conv2) at (0, -7) {5x5 Conv, 16 Kernels};
\node[bn] (bn2) at (0, -8.5) {Batch Normalization};

\node[pool] (pool2) at (0, -11) {2x2 Max Pooling};

\node[conv] (conv3) at (0, -12.5) {3x3 Conv, 32 Kernels};
\node[bn] (bn3) at (0, -14) {Batch Normalization};

\node[pool] (pool3) at (0, -16.5) {2x2 Max Pooling};

\draw[->] (input.south) -- (conv1.north);
\draw[->] (conv1.south) -- (bn1.north);
\draw[->] (bn1.south) -- (pool1.north);

\draw[->] (pool1.south) -- (conv2.north);
\draw[->] (conv2.south) -- (bn2.north);
\draw[->] (bn2.south) -- (pool2.north);

\draw[->] (pool2.south) -- (conv3.north);
\draw[->] (conv3.south) -- (bn3.north);
\draw[->] (bn3.south) -- (pool3.north);

\draw[decorate,decoration={brace,amplitude=10pt,raise=4pt},yshift=0pt]
  (3.25, -1.2) -- (3.25, -3.3) node[midway,xshift=40pt,font=\Large] {4 times};

\draw[decorate,decoration={brace,amplitude=10pt,raise=4pt},yshift=0pt]
  (3.25, -6.7) -- (3.25, -8.8) node[midway,xshift=40pt,font=\Large] {4 times};

\draw[decorate,decoration={brace,amplitude=10pt,raise=4pt},yshift=0pt]
  (3.25, -12.2) -- (3.25, -14.3) node[midway,xshift=40pt,font=\Large] {4 times};

\end{scope} 

\end{tikzpicture}}
    \caption{Architecture of the CNN, the left figure (a) an overview is given and in the right figure (b) the full convolutional block is shown.}
    \label{fig:overview architecture cnn}
\end{figure}
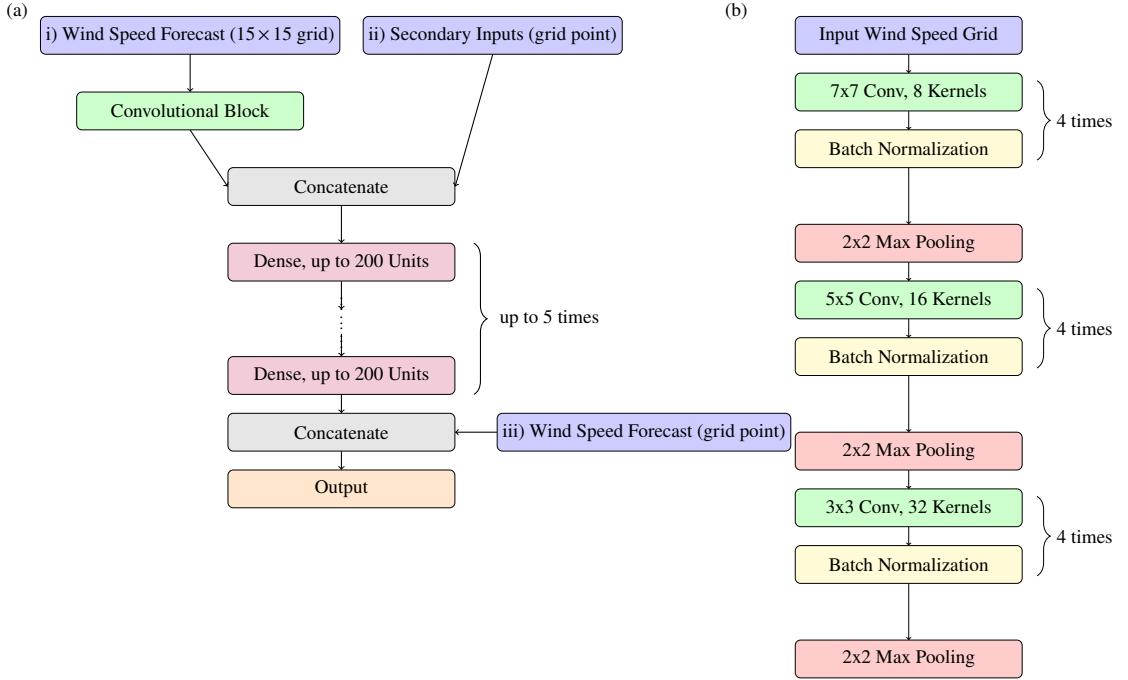

The input of the CNNs is split into three different parts, as shown in Figure \ref{fig:overview architecture cnn}. The first part is a grid from the HA40 wind speed forecasts with the weather station located in the central grid point (in Figure \ref{fig:overview architecture cnn}a.i). This part of the input is not normalized and is handled by the convolutional part of the neural network to learn spatial features. The second part of the input contains the variables from the HA40 forecasts in Table \ref{tab:predictors}, except for the wind speed at 10 meter height. These other four variables are called the \textit{secondary input} and are shown in Figure \ref{fig:overview architecture cnn}a.ii). 
 We chose an architecture similar to the one used by \citet{Daniel}, in which the grid input is handled by a convolutional block and then concatenated with the secondary inputs. This is then passed to a number of dense layers. The convolutional block is specified in \ref{fig:overview architecture cnn}b.
 The third part of the input is the grid point forecast of HA40 wind speed. This input is passed to the final dense layer, before the output of the network is generated (Figure \ref{fig:overview architecture cnn}a.iii). This is added for similar reasons to the ones used by \citet{Daniel}, in which it is argued that there is a close to linear relation between the HA40 forecasts and the observations. By passing the HA40 wind speed forecast directly to the final dense layer, the CNN can better focus on modeling the deviation from this linear relation. 

\subsection{Parametric Distribution}\label{sec:parametric distribution}
 
A popular distribution for wind speeds is the truncated normal (TN) distribution \citep{Thorarinsdottir_ProbForecastsWindSpeed_2010}. Given that the main aim is to improve the prediction of extremes, a more heavy right-tailed distribution could be better because these distributions assign a larger probability to higher wind speeds. Two examples of such distributions that have been used for wind speeds are the log-normal (LN) distribution \citep{LogNormalEMOS} and the generalized extreme value (GEV) distribution \citep{Lerch_2013_GEV, Baran_2021_trunc_gev}. We use these same three distributions, where we use a similar strategy to \citet{Thorarinsdottir_ProbForecastsWindSpeed_2010} for the TN and \citet{LogNormalEMOS} for LN distributions, using \eqref{eq:emos mean ensembles} and \eqref{eq:emos var features} to model the location and scale parameters. For the GEV distribution the implementation is similar to \citet{Lerch_2013_GEV}, where we again use \eqref{eq:emos mean ensembles} and \eqref{eq:emos var features} to model the location and scale parameters, and use a free shape parameter.

It is also possible to combine multiple distributions to both use the capabilities of the TN for the forecast of bulk observations and LN or GEV distribution for more extreme forecasts. \citet{MixtureEMOS} first proposed this method, where a regime switching distribution is employed. The regime-switching approach consists of using a specific threshold $t \in \R$ to determine which distribution to use: if the predicted wind speed from the NWP model is smaller than $t$, the forecast distribution is a TN, otherwise it is a heavier-tailed distribution, such as the GEV or LN. 

\citet{MixtureEMOS} also suggest to use mixture distributions. In the simplest case, a mixture distribution is defined using two cumulative distribution functions $F_1, F_2$ and a weight $w \in [0,1]$, where the cumulative distribution function of the mixture takes the form
\begin{align*}
    F(x) = w F_1(x) + (1 - w) F_2(x), \qquad x\in \R.
\end{align*}
During training, the weight of the mixture distribution is estimated \citep{MixtureEMOS}. However, in this simple case, the weight is independent of the inputs. Allowing the weight to depend on the weather situation may be helpful to obtain superior forecast performance. For example, similar to the regime switching distribution,  it may be beneficial to use one distribution for modeling higher wind speeds and another distribution for modeling lower wind speeds. We make our weight parameter dependent on the input using the model
\begin{equation}
\label{eq:adaptive weight parameter}
    w(X_w) = \text{sigmoid}(\alpha + \beta X_w) = \frac{1}{1+e^{-(\alpha + \beta X_w)}},
\end{equation}
 where $X_w$ is the NWP forecast of the wind speed and $\alpha,\beta\in \R$ are two parameters to fit using training data. Assuming that $X_w$ is an appropriate proxy for the observed wind speed, it provides a simple way to determine whether a heavier right-tailed distribution is needed. We refer to this type of distribution as an \textit{adaptive mixture} (AM) distribution.

Regarding the choice of the sigmoid function, it meets the criteria of being a continuous, strictly increasing function with a range in the interval $[0,1]$. Alternatively, other functions, such as the standard Gaussian cumulative distribution function, could also be considered. Another approach is to use additional features to determine the weight parameter. However, we did not explore further experimentation with the Gaussian CDF, alternative functions, or the inclusion of additional features.

\subsection{Scoring Rules}

Next to the choice of parametric distribution, we are also interested in the effect of different scoring rules as loss function. Traditionally, the parameters within EMOS are optimized using the negative log-likelihood or the continuous ranked probability score (CRPS), which are both proper scoring rules \citep{GneitingCRPS}. The CRPS is defined by \begin{align}\label{eq:CRPS}
\begin{split}
    \text{CRPS}(F,y) &= \int^\infty_{-\infty}(F(z) - \1\{y\leq z\})^2 \text{d}z \\
    &= \E_F(|X-y|) - \half \E_F(|X-X'|),
\end{split}
\end{align}
where $X,X'$ are independent random variables with distribution $F$ \citep{GneitingCRPS}. We will test the effect of training using the \textit{weighted CRPS (wCRPS)} \citep{Gneiting_twCRPS_2011}, a generalization of the CRPS that uses a weighting function $w:\R \rightarrow \R$ to emphasize important regions in the forecast. The wCRPS is defined by
\begin{align}\label{eq:twCRPS}
\begin{split}
    \text{wCRPS}(F, y;w) &= \int^\infty_{-\infty}(F(z) - \1\{y\leq z\})^2 w(z)\text{d}z \\
    &= \E_F(|v(X) - v(y)|) - \half \E_F(|v(X) - v(X')|),
\end{split}
\end{align}
where $X,X'\sim F$ are independent and $v$ is any function such that $v(z) - v(z') = \int^z_{z'} w(x) \text{d}x$, called the \textit{chaining function} \citep{Taillardat_evaluating_prob_forecast_of_extremes_using_crps_distributions_2023}. The proof of the second equality in \eqref{eq:twCRPS} is given in Proposition 1 of \cite{Allen2022EvaluatingKernelScores}. An important property of the wCRPS is that it is a proper scoring rule. The final expression in \eqref{eq:twCRPS} is the starting point for a sampling-based approach to approximate the wCRPS. We use this appoach as there are only few weight functions for which there is a closed-form expression for the wCRPS.

We consider two types of weight functions. First of all, we consider the indicator function 
\begin{align}\label{eq:weight function indicator}
    w(z) = \1_{\{t\leq z\}}
\end{align}
for some threshold $t\in \R$, with chaining function $v(z) = \max\{z, t\}$. This is an interesting weight function because it provides insight in the performance of the forecast above the threshold $t$. When this weight function is used, then the associated scoring rule will be called the \textit{threshold-weighted continuous ranked probability score (twCRPS)}. Although this name is also used in the literature for the wCRPS, we will make a distinction to prevent any confusion.
Some examples of where the twCRPS is applied for the evaluation of extreme events are \cite{Lerch_2013_GEV, LogNormalEMOS, AllenReferencetwCRPS, MixtureEMOS}. An advantage of the twCRPS is that closed-form expressions are available for several distributions \citep{wessel2024improvingprobabilisticforecastsextreme}.

The second type of weight function is 
\begin{align}\label{eq:weight normal cdf constant}
    w(z) = c + \Phi_{\mu,\sigma}(z),
\end{align}
where $c \in\R$ and $\Phi_{\mu,\sigma}:\R \rightarrow \R$ is the cumulative distribution function of a normal random variable with mean $\mu$ and standard deviation $\sigma$. By taking the derivative of this weight function one can find the associated chaining function
\begin{align}
    v(z) = cz + (z - \mu)\Phi_{\mu,\sigma}(z) + \sigma^2 \phi_{\mu,\sigma}(z),
\end{align}
where $\phi_{\mu,\sigma}:\R\rightarrow \R$ is the probability density function of a normal random variable with mean $\mu$ and standard deviation $\sigma$.  

The weight function in \eqref{eq:weight normal cdf constant} is useful because it gives the user control over the weight assigned to lower wind speeds compared to higher wind speeds in a flexible fashion. This will prove to be useful when we want to produce forecasts that mainly take into account the performance on high wind speeds, while not ignoring the performance on lower wind speeds, which can be achieved by changing the value of $c$.

\subsection{Implementation}

We will now describe the implementation details of the experiments. All models are trained globally, meaning that a single model is used for all locations, instead of fitting a unique model for each station. 
A sampling-based approach is employed to approximate the loss function values in \eqref{eq:CRPS} and \eqref{eq:twCRPS}. Concretely, in case a sample size of $n$ is used with a chaining function $v:\R \rightarrow \R$, the wCRPS is estimated by
\begin{align}\label{eq:wCRPS sampling based estimate}
    \frac{1}{n} \sum^n_{i=1}\left(|v(x_i) - v(y)| - \half |v(x_i) - v(x_i')|\right),
\end{align} 
where $x_1,\ldots,x_n,x_1',\ldots,x_n'$ are independently drawn samples from the forecast distribution and $y$ is the observation. In this approach, we draw a new set of samples for each forecast and observation in our training set. To estimate the CRPS with a sampling-based approach, we use \eqref{eq:wCRPS sampling based estimate} with $v(z)=z$. In Appendix C we compare the use of a sampling-based approximation of the CRPS to the use of a closed-form expression and observe that both yield a similar model performance. This strongly suggests that using a sampling-based approximation of the wCRPS is a reasonable approach.

To optimize the parameters in our model, we need to be able to obtain the gradients of \eqref{eq:wCRPS sampling based estimate} with respect to parameters of a probability distribution. For this purpose we use the well-known the reparametrization trick (see, e.g., \cite{Goodfellow}), i.e., we represent a random variable $z$ as a function of a random variable $\epsilon$ and the distribution parameters $\theta$, i.e., $z=f(\epsilon, \theta)$, where $\epsilon$ follows a fixed distribution. 
For example, if $z \sim \mathcal{N}(\mu,\sigma^2)$ we can write $z = \mu + \sigma \epsilon$, where $\epsilon \sim \mathcal{N}(0,1)$. For the TN, LN, and GEV distributions one can proceed similarly. 

EMOS with a linear parameter model is implemented in TensorFlow \citep{tensorflow2015-whitepaper}, with the help of the TensorFlow Probability extension. Additionally, for the CNNs we use Keras \citep{chollet2015keras} as frontend. TensorFlow inherently supports the reparametrization trick within its framework, allowing for efficient gradient computation with random variables. The full code for this project can be found on Github, see \citet{GithubSimon}.

\subsubsection{Implementation Linear Models}
For EMOS with linear regression, the input consists of the features from Table \ref{tab:predictors}. For all the features we subtract the training data mean and divide by the training data standard deviation.

We implement all parametric distributions from section 3\ref{sec:parametric distribution}, where $\mu$ and $\sigma$ are computed as in \eqref{eq:emos mean ensembles} and \eqref{eq:emos var features}. We use a fixed initialization for the coefficient vectors $a, b\in \R^{n+1}$, setting all their entries to 1.
For all distribution we use $\mu$ and $\sigma$ as the location and scale parameters of the distribution. For the GEV distribution, we use a shape parameter that is fitted to the data as a free parameter during training. Once fitted, the shape parameter remains constant. During optimization, there are two possible settings for the shape parameter. When $\xi_G$ is initialized to zero, the distribution would be and remain of the Gumbel type. In case $\xi_G$ is set to any value other than zero, the distribution can switch between the Fréchet and Gumbel type. During validation we found that the second case led to better results, thus we initialize $\xi_G$ to $0.3$. Every initialization for $\xi_G$ leads to convergence, with values between $-0.5$ and $0.5$ leading to faster convergence, thus our choice for 0.3.

For the mixture models, we have three sets of parameters. First, we have $\mu_1$ and $\sigma_1$, which are determined by \eqref{eq:emos mean ensembles} and \eqref{eq:emos var features} respectively. They are used to model the location and scale parameters of the first distribution. Similarly, we have $\mu_2$ and $\sigma_2$, which are used to model the location and scale parameters of the second distribution. The associated parameters $a,c,a',c' \in \R^{n+1}$ are optimized simultaneously. Finally, we also have a weight parameter. In case of the mixture distribution, this is a parameter in $[0,1]$, initialized at $0.5$, and its constraint is enforced by projecting back onto $[0,1]$ after each iteration. For the adaptive mixture distribution, we need to optimize the parameters $\alpha, \beta \in \R$ in \eqref{eq:adaptive weight parameter}. The idea is that the heavier-tailed distribution, such as the LN or GEV should be used for the higher wind speeds and the TN for lower wind speeds. During training, it was found that this was not always the outcome. Consequently, we place additional constraints on $\alpha$ and $\beta$. We choose these constraints to be $\alpha \in [4,12]$ and $\beta \in [-6, -0.6]$, and initialize them at $\alpha=5$ and $\beta=-1$. Through visual inspection of the weight function, we found that these ranges ensure that the first distribution is used for the lower wind speeds and the second distribution for higher wind speeds. The initialization of $\alpha$ and $\beta$ is set to these values because they satisfy the constraints. The constraints are enforced by projecting $\alpha$ and $\beta$ back into their respective ranges after every iteration.

For the models that have a mixture distribution, we utilize an approach to decrease training times: we first pre-train each single parametric distribution for 75 epochs, then use these to initialize training of the mixture model. Results have been checked to ensure that this does not change the performance of the models.
For the linear models, there is no problem with overfitting, which is checked by monitoring the validation loss during cross-validation. Thus, all models are trained until the training loss converged.

We also experimented with random initialization, in which the parameters are initialized to a sample from a standard Gaussian distribution. This does not lead to better results, hence we use fixed initialization since this produces more consistent results. Alternative distributions for initialization are not considered.

For training, we consider sampling-based approximations of the CRPS and wCRPS as loss functions. For the wCRPS, we experiment with two possible weight functions. The first one is the indicator function from \eqref{eq:weight function indicator}, and the second weight function is the shifted Gaussian CDF from \eqref{eq:weight normal cdf constant}. In the sampling-based approach a sample size of 250 is used. The training was also observed to be stable for smaller sample sizes, such as 100. The value of 250 is selected to ensure we have consistent and fast convergence.

Next to the hyperparameters in the weight function of the wCRPS and the choice of parametric distribution, there are more hyperparameters that need to be selected. These hyperparameters are the batch size, the optimizer (Adam or SGD), and the learning rate of the optimizer. These hyperparameters are selected using cross-validation. In section 3\ref{sec:model selection}, we will further elaborate on the validation loss and the algorithm used for hyperparameter tuning.

\subsubsection{Implementation CNN Models}

As can be seen in Figure \ref{fig:overview architecture cnn}, the input of the CNN models consists of several parts. For the secondary input, we subtract the training data mean and divide by the training data standard deviation. This normalization procedure is also applied to the test data, utilizing the mean and standard deviation of the training data.

The output of the neural network depends on the probability distribution that is chosen. For the TN, the output is the location and the scale of the normal distribution. The activation function for the mean is linear, and for the variance we use the softplus function. For the LN it contains the mean and variance of the underlying normal distribution, again using the linear and softplus activation function.  In case a mixture distribution of TN and LN is used, the output consists of two sets of parameters of the respective distributions and a weight parameter, with a sigmoid activation function. Note that this can be seen as an extension of the adaptive mixture model. Throughout the rest of the network, the ReLU activation function is used. We also implemented the GEV distribution, where the output of the neural network contains the location, scale, and shape parameters of the distribution. However, training the network with the GEV distribution turned out to be very unstable. Hence we did no further experimentation with the GEV distribution in combination with CNNs.

 As with the linear models, the models are trained on sampling-based estimates of the CRPS and wCRPS, using equation \eqref{eq:wCRPS sampling based estimate}. In the sampling-based approach a sample size of 1000 is used. We choose a larger sample size compared to the linear model, since the convergence of CNNs is less stable than for the linear models. For the CNN-based model we choose the largest sample size that still resulted in reasonable training times.

During hyperparameter tuning, we tune the parameters of the weight function in the wCRPS, the choice of distribution, as well as several other parameters: the optimizer (Adam or SGD), the learning rate of the optimizer, the batch size, the number of dense layers, the number of units per dense layer and the hyperparameter of the $\ell_2$ penalty for the dense layers.

We observed that overfitting is a problem when training CNNs. Accordingly, during validation we utilize early stopping, with a patience of 10 epochs. When training the final models, we use two-thirds of the epochs that yield the best results during validation, since the full training data set is used to train the final models. This `best number of epochs' is obtained by averaging the number of epochs that is deemed optimal on each fold, by training 30 models on each of the three folds. 

\subsection{Model Selection}\label{sec:model selection}

We use two different metrics to test the performance of a model during hyperparameter tuning. The first one is the CRPS, which provides an overall performance measure over the entire range of wind speeds. The second scoring metric is the twCRPS, so that we can assess model performance on the extremes. We choose an indicator weight function at threshold 12 m/s (twCRPS12), which is approximately the 95$^{th}$ percentile of the training data (Table \ref{tab:data quantiles}).

During validation, we observed that the CRPS and twCRPS12 are competing objectives. Therefore, one can only hope to identify a Pareto front containing the best trade-offs. We use the \textit{multiobjective tree-structured parzen estimator} (MOTPE) \citep{MOTPE} for this task. MOTPE is a hyperparameter optimization algorithm that efficiently balances competing objectives by modelling their distributions, enabling the identification of Pareto-optimal solutions that represent the best trade-offs.

The MOTPE implementation from Optuna by \citet{optuna_2019} is used, starting every search with 20 random trials to get a general impression of the search space and model the distribution of the objectives. We then performed around 200 trials with MOTPE for both the linear and CNN models. 

\section{Results}\label{sec:results}

The results are divided into three different subsections. First, we look at the results for the models using linear regression. Then, the results for the CNN-based models will be presented. Finally, we examine the effect of transferring a Pareto-optimal choice of weight function for the wCRPS for the linear models to a CNN-based model, and from the CNN-based model to the linear models. 

\begin{figure}[h]
    \centering
\includegraphics[width=0.9\linewidth]{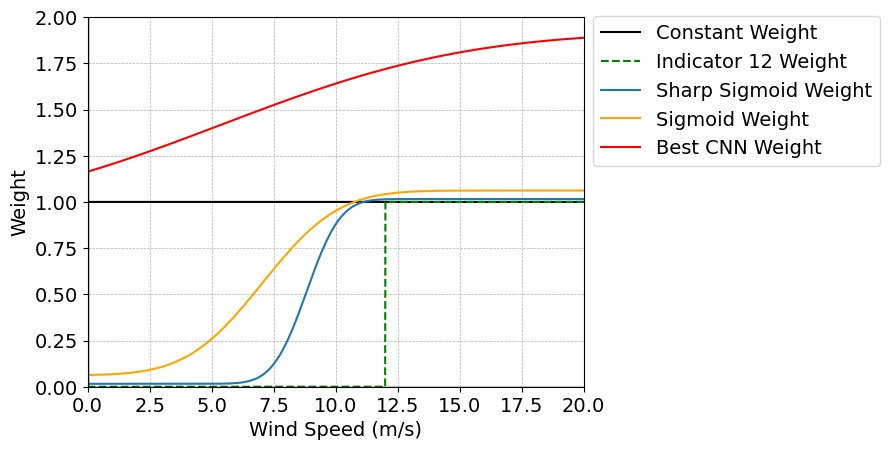}
    \caption{Weight functions for the models from Table \protect{\ref{tab:hyperparameters weight function}}.}
    \label{fig:weight functions plotted}
\end{figure}

In Table \ref{tab:hyperparameters weight function} and in Figure \ref{fig:weight functions plotted} we summarize all used weight functions together with a short description (under `Model'). The constant weight leads to the CRPS and is used to train the standard EMOS as a reference model. The `indicator weight' refers to the indicator function \eqref{eq:weight function indicator} with threshold 12 m/s. The `sharp sigmoid weight' and `sigmoid weight' are weight functions of the form \eqref{eq:weight normal cdf constant} with hyperparameters selected from the Pareto front for the linear models. Finally, the hyperparameters of `best CNN weight' are selected from the Pareto front for the CNN-based models. 

\begin{table}[h]
\centering
\begin{tabular}{lcccccc}
\toprule
\textbf{Model} & \textbf{Loss Function} & \textbf{$\bm{\mu}$} & \textbf{$\bm{\sigma}$} & \textbf{$\bm{c}$} & \textbf{$\bm{t}$} & \textbf{Model Type} \\
\midrule
Constant Weight          & CRPS          & -   & -  & - &- & Reference Model \\
\midrule
Indicator Weight & twCRPS & - & - & - & 12 \\
Sharp Sigmoid Weight    &  wCRPS        & 8.84   & 1.07  & 0.02&- & Linear Model \\
 Sigmoid Weight      & wCRPS          & 7.05   & 2.41  & 0.06&-&  \\
\midrule
Best CNN Weight & wCRPS & 5.42 & 7.82 & 0.92 & -& CNN Model \\
\bottomrule
\end{tabular}
\caption{Selected hyperparameters of the different models. The values $\mu,\sigma $ and $c$ correspond to the parameters from the weighting function from equation \protect{\eqref{eq:weight normal cdf constant}}. Model type refers to the model for which the weight function is selected.}
\label{tab:hyperparameters weight function}
\end{table}

\subsection{Results for Linear Models}\label{sec:results linear emos}

 The hyperparameters of all models are chosen from the Pareto front identified during the hyperparameter optimization of the linear models, with the selected values shown in Table \ref{tab:linmodelhyperparameters}. The choice of optimizer and learning rate did not change the results, hence in all models we use Adam with a learning rate of 0.01. The same holds for the batch size, any choice larger than 32 gives similar results, so we fixed it at 256. We chose this value because this results in the fastest training times, although the improvement was minor.

\begin{table}[h]
    \centering
    \begin{tabular}{lc}
    \toprule
    \textbf{Hyperparameter} & \textbf{Value} \\
    \midrule
    Optimizer               & Adam           \\
    Learning Rate           & 0.01       \\
    Batch Size              & 256             \\
    \bottomrule
    \end{tabular}
    \caption{Hyperparameters for linear models.}
    \label{tab:linmodelhyperparameters}
\end{table}

\begin{figure}
    \centering
    \includegraphics[width=0.6\linewidth]{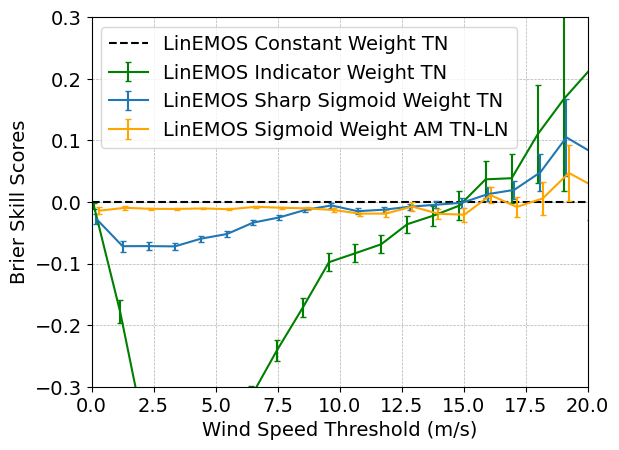}
    \caption{Bootstrapped Brier skill scores on the test set, with a bootstrap size of 10.000.}
    \label{fig:bootstrap bss linear models test set}
\end{figure}

In Figure \ref{fig:bootstrap bss linear models test set} we can see the bootstrapped Brier skill scores for different model configurations, using the linear model with constant weight (i.e., the CRPS) and TN as parametric distribution as the reference model. 
We see that the TN distribution combined with an indicator weight function performs very well above 15 m/s, although it also exhibits very bad performance below this threshold, where the BSS drops to around $-0.5$ (green line in Figure \ref{fig:bootstrap bss linear models test set}). For the model with a sigmoid weight function and TN distribution, this worsening in performance for moderate wind speeds is dampened, while this also limits the increase in BSS for higher wind speeds. The model with the sigmoid weight has the adaptive mixture (AM) distribution as parametric distribution (yellow line in Figure \ref{fig:bootstrap bss linear models test set}). The hyperparameters of this model were obtained from the Pareto front with a larger focus on moderate wind speed, compared to the other models. However, this model closely resembles the constant weight with TN in terms of BSS. 
 We can also see that the variance from bootstrapping becomes very large above 15 m/s. The result of improved performance in the extreme wind speed range at the cost of worse performance on moderate wind speeds is in line with the results found by \citet{wessel2024improvingprobabilisticforecastsextreme}. 

\begin{figure}[h]
    \centering
    \includegraphics[width=0.9\linewidth]{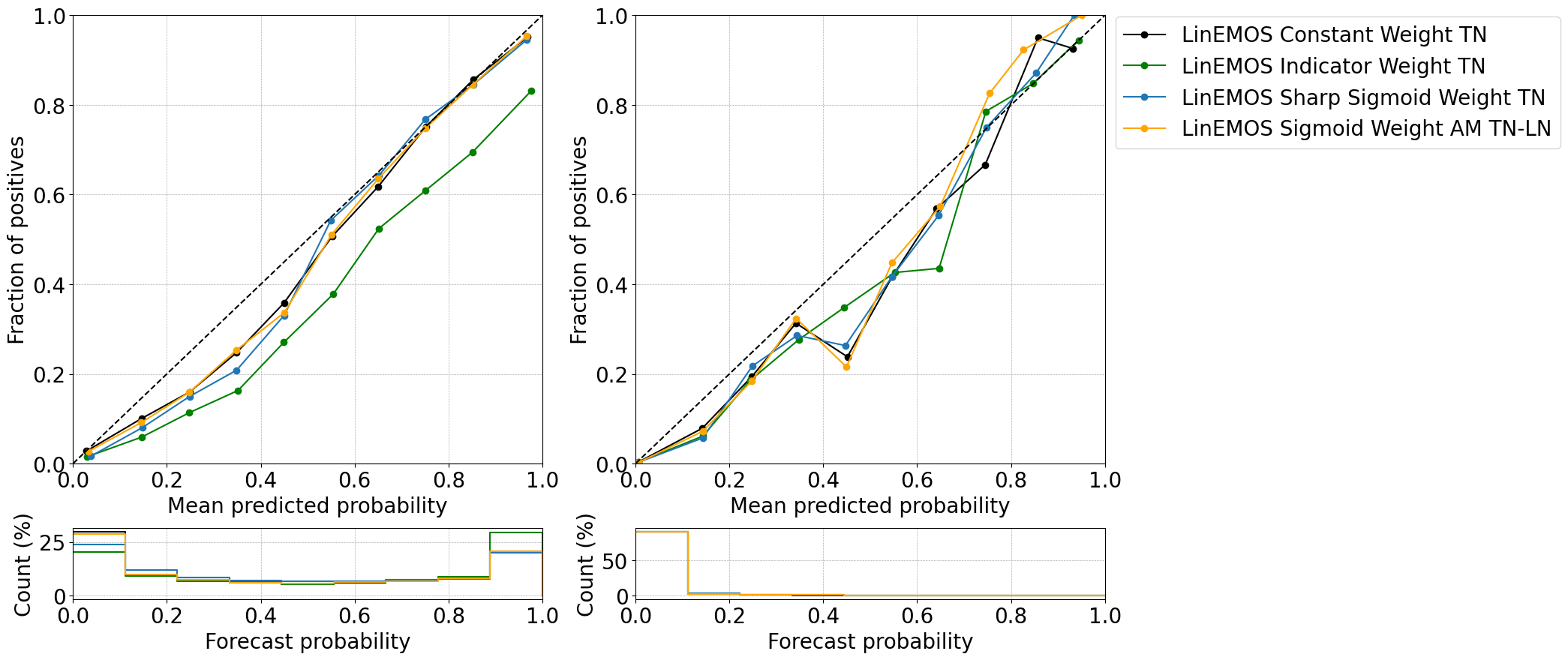}
    \caption{Reliability and sharpness diagrams at a threshold of 5 m/s (left figure) and 12 m/s (right figure).}
    \label{fig:reliability linear models}
\end{figure}

Reliability diagrams show that the model with the indicator weight is not well calibrated for forecasts of wind speed exceeding 5 m/s, as it has a positive bias (Figure \ref{fig:reliability linear models}). A similar, although less pronounced, pattern can be observed for the other models. All models are well calibrated for the probability of wind speeds exceeding 12 m/s, and there are few significant differences between the models. Also the sharpness is almost identical at this threshold (Figure \ref{fig:reliability linear models}).

\begin{figure}[h]
        \centering
        \includegraphics[width=0.5\textwidth]{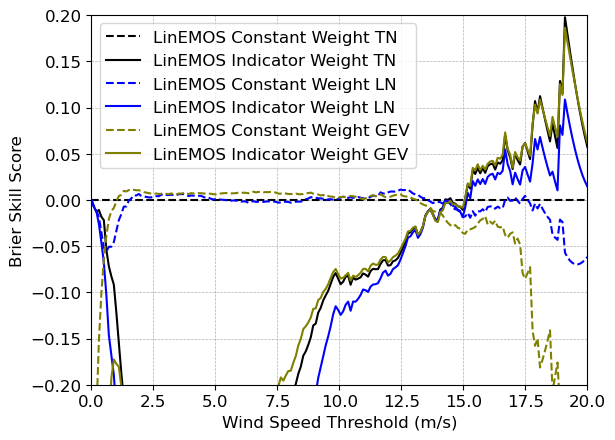} 
        \caption{Brier skill scores on the test set with different distributions. The solid lines lie below the range of the figure between 1 and 8 m/s.}
        \label{fig:linear models different distribution}
\end{figure}

Finally, we look at the effect of choosing a different parametric distribution (TN, LN, GEV), when optimizing the constant and indicator weight functions (Figure \ref{fig:linear models different distribution}). With a constant weight function, the LN and GEV distributions are slightly better between 1 and 12 m/s, with worse performance for extremes. The GEV model performs very poorly below 1 m/s, presumably because it can predict negative wind speeds, something that does not occur for the LN and TN. When we train on the indicator function, all distributions perform poorly for lower wind speeds. Above 15 m/s, the TN, LN and GEV combined with the indicator weight all have very similar BSS.

\subsection{Results for CNNs}

\begin{figure}
    \centering
    \includegraphics[width=0.9\linewidth]{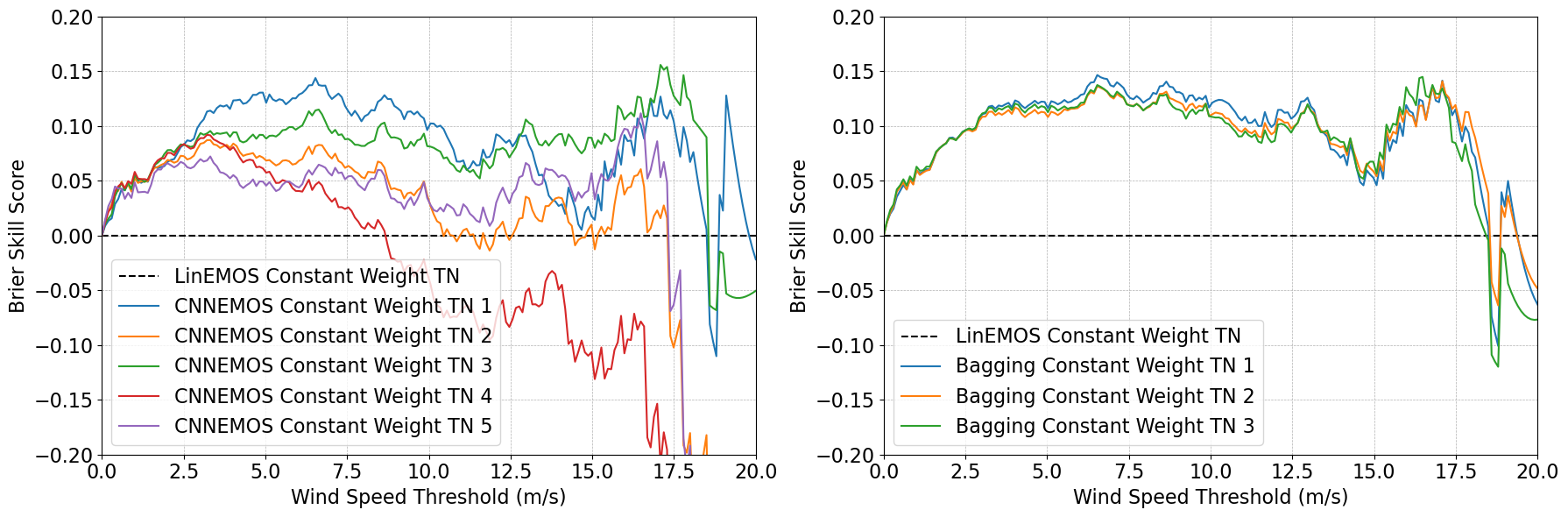}
    \caption{Brier skill scores (compared to linear models with a constant weight function and a TN distribution) of 5 single CNNs (left) and 3 bagging CNNs (right). All models are trained with the exact same hyperparameters.}
    \label{fig:variability cnns}
\end{figure}

A significant amount of variability in performance was observed between different training runs of the CNN models. This stems from three different sources of this randomness; batch selection in optimization, the sampling-based approach in estimating the CRPS and wCRPS, and the random parameter initialization at the start of training. As a consequence,  comparing single CNNs can be misleading. This variability of different runs is showcased in the left display in Figure~\ref{fig:variability cnns}. As a solution, we used a `bagging' version of CNNs, where we trained 10 models with the exact same hyperparameters, and then combined the 10 different predictive distributions into a single distribution by taking their average. The resulting distribution is then used to compute the different verification scores. Three different runs of the bagging model are shown in the right display in Figure~\ref{fig:variability cnns}. In Figure~\ref{fig:forecasts pdf bagging 2 figures} in Appendix A the pdfs of the bagging estimator are shown for two different samples from the test data. In certain scenarios, the individual models all agree on what the distribution parameters are (left hand side of Figure~\ref{fig:forecasts pdf bagging 2 figures}), while in other cases their predictions vary (right hand side of Figure~\ref{fig:forecasts pdf bagging 2 figures}). 

For hyperparameter optimization, we employed a two-step approach. In the first step, we used a broader search space, to optimize the parameters listed in Table \ref{tab:cnnhyperparameters}, along with the weight function and the parametric distribution. The search space is shown in Table \ref{tab:searchranges cnn hypreopt} in Appendix A. During this phase, the hyperparameters in Table \ref{tab:cnnhyperparameters} were much more influential than the choice of weight function and distribution. Therefore, after this first round of optimization, we fixed the found values for the hyperparameters in Table \ref{tab:cnnhyperparameters} and then performed another round of hyperparameter optimization, focusing solely on the weight function and distribution. 

To train the final models, the number of epochs during training was varied per model, since certain combinations of weight functions and distribution required more epochs to train during validation. These values can be found in Table \ref{tab:modelepochs}. 
The Bagging Constant Weight TN model serves as a reference model. The hyperparameters of the Best CNN Weight function were selected by the MOTPE algorithm as one of the Pareto optimal solutions. Both the TN and the mixture distribution of TN and LN were frequently selected as a good distribution and therefore we included both distributions in the final models. Since the LN distribution was not selected in the best configurations of the MOTPE algorithm, we did not train any final model with this distribution.

\begin{table}[h!]
    \centering
    \begin{tabular}{lc}
        \toprule
        \textbf{Hyperparameter} & \textbf{Value} \\
        \midrule
        Optimizer               & Adam           \\
        Learning Rate           & 0.000105       \\
        Batch Size              & 64             \\
        Dense Layers            & 2              \\
        Units per Dense Layer   & 170            \\
        $\ell_2$ Penalty           & 0.031658      \\
        \bottomrule
    \end{tabular}
    \caption{Hyperparameters for CNNs.}
    \label{tab:cnnhyperparameters}
\end{table}

\begin{table}[h!]
    \centering
    \begin{tabular}{lc}
        \toprule
        \textbf{Model Name} & \textbf{Epochs} \\
        \midrule
        Bagging Constant Weight TN          & 47 \\
        Bagging Best CNN Weight TN          & 47 \\
        Bagging Best CNN Weight TN-LN       & 50 \\
        \bottomrule
    \end{tabular}
    \caption{Number of epochs used to train the different models.}
    \label{tab:modelepochs}
\end{table}

\begin{figure}
    \centering
    \includegraphics[width=0.5\linewidth]{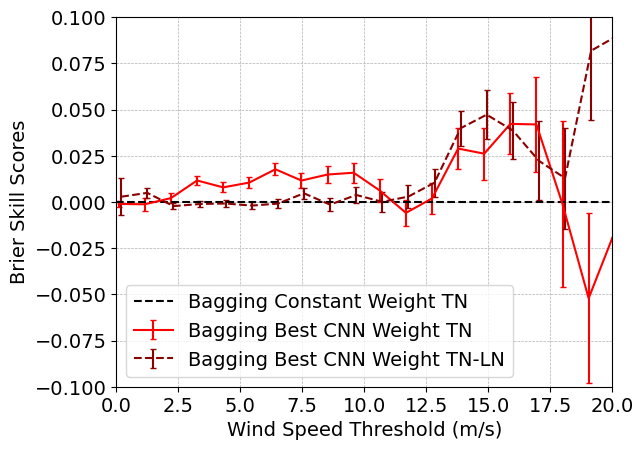}
    \caption{Bootstrapped Brier skill scores on the test set, with a bootstrap size of 10.000 and the bagged TN model trained on the CRPS used as reference}
    \label{fig:bootstrap bss bagging models}
\end{figure}

There is no noticeable body-tail trade-off for the best CNN weight function, since the BSSs are approximately equal until 12 m/s (Figure \ref{fig:bootstrap bss bagging models}). Above this threshold, the best CNN weight function with the mixed TN-LN distribution improves upon the constant weight function  with the TN. 

\begin{figure}
    \centering
    \includegraphics[width=0.9\linewidth]{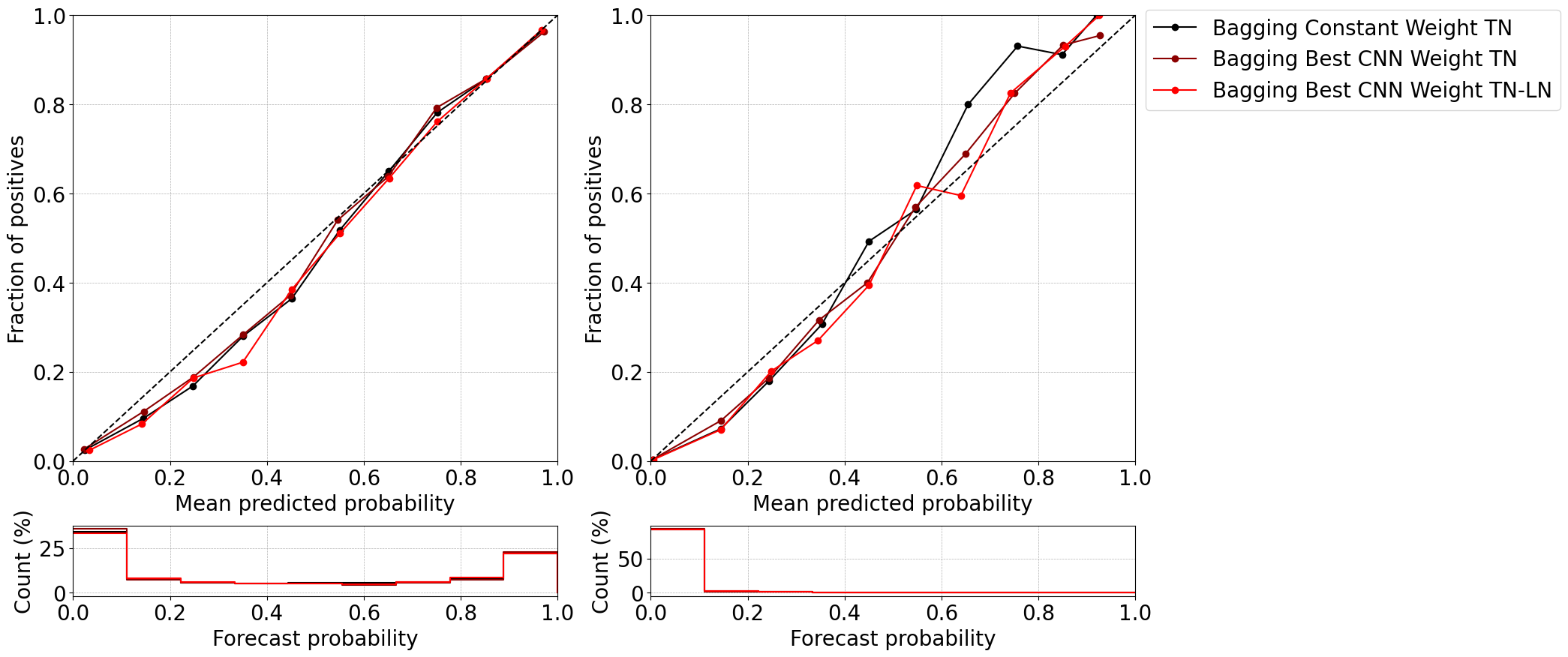}
    \caption{Reliability and sharpness diagrams at a threshold of 5 m/s (left figure) and 12 m/s (right figure).}
    \label{fig:reliability bagging models}
\end{figure}

In terms of calibration, all models perform equally well. This can be seen in Figure \ref{fig:reliability bagging models}, which shows the reliability and sharpness diagrams at 5 and at 12 m/s. 
In line with the linear models, there is not a big difference in performance between the models using a TN and the model using a TN-LN mixture distribution. 

\begin{figure}
    \centering
    \includegraphics[width=0.6\linewidth]{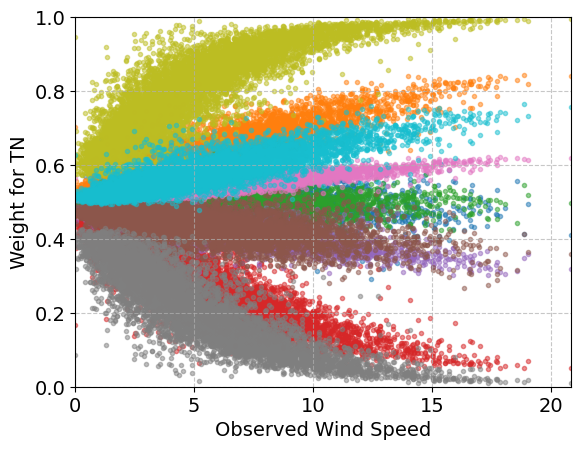}
    \caption{Weight for the TN distribution in a TN-LN mixture distribution for all 10 CNN models from a bagging model.}
    \label{fig:weight all bagging models}
\end{figure}

For the linear models with the adaptive mixture distribution a constraint was used since the models were not consistently choosing the TN for the lower wind speeds and the GEV or LN for higher wind speeds. For the mixture distributions with CNNs we had a similar result. In Figure \ref{fig:weight all bagging models} we can see how much weight is assigned to the TN compared to the LN on the test set. We see strongly varying results, where in certain models only the TN or LN is used for extremes, and in other models the weight is close to a half for all the samples in the test set.

\begin{figure}
    \centering
    \includegraphics[width=0.5\linewidth]{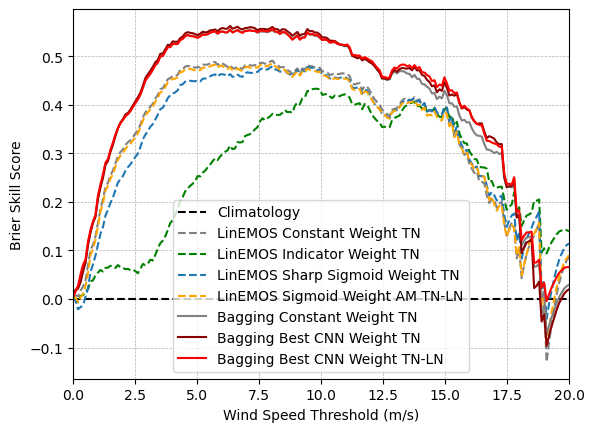}
    \caption{Brier skill scores on the test set with climatology as reference model.}
    \label{fig:bss climatology all models}
\end{figure}

Finally, Figure \ref{fig:bss climatology all models} contains the BSS for both the linear and the bagged CNN models with climatology as reference. Climatology is the probabilistic forecast based solely on past measurements, and a separate climatology model is constructed for each observation station. Between 0 and 12 m/s the bagged CNN models are the best performers regardless of the weight function. 
Between 12 and 17 m/s the bagged CNN models with the best CNN weight have the best results, and above 17 m/s the linear EMOS model with indicator weight has the best BSS, although the skill for this model at moderate wind speed values is much lower than the other post-processed models. 

\subsection{Transferring Weight Function}\label{sec:transferring weight function}

\begin{figure}
    \centering
    \includegraphics[width=0.9\linewidth]{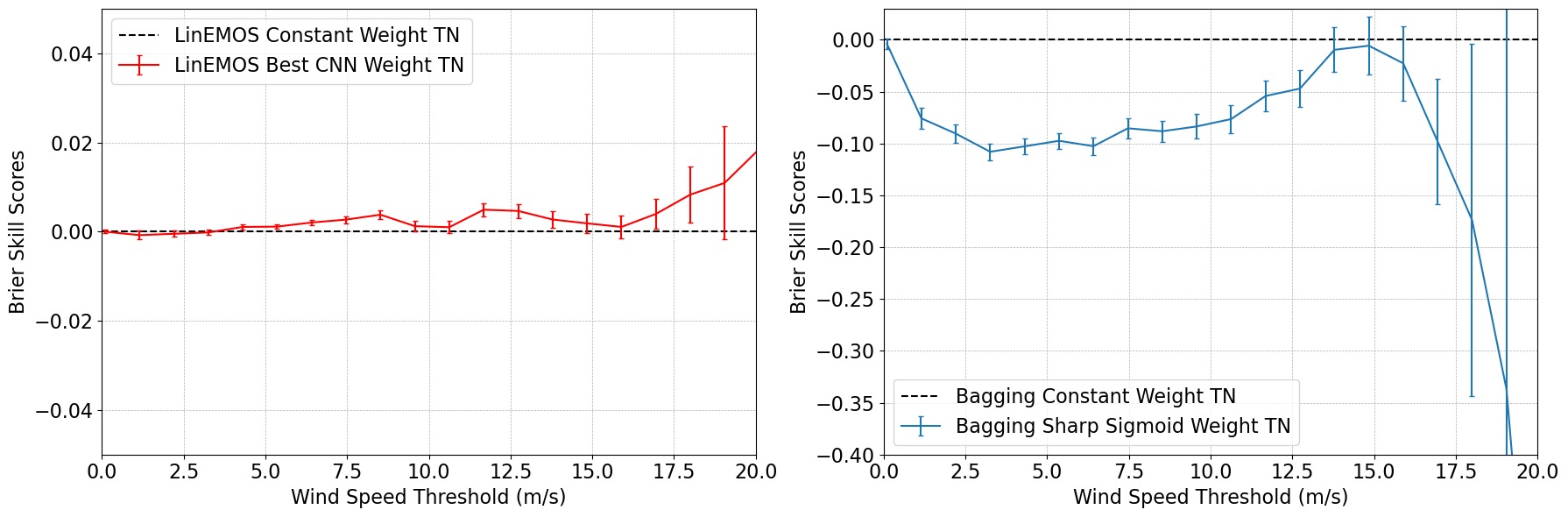}
    \caption{Bootstrapped Brier skill scores for a linear model using the Best CNN Weight (left) and a bagging model using the Sharp Sigmoid Weight (right).}
    \label{fig:bootstrap bss transfer weight}
\end{figure}

We will now look specifically at the effect of transferring a weight function that yields skillful forecasts on extremes for one model to another model. We train a CNN-based model with the sharp sigmoid weight function that performed well on extremes for linear models and, conversely, we also train a linear model with the best CNN weight function. In Figure~\ref{fig:bootstrap bss transfer weight} (left) we can see the BSS of a linear EMOS model with the best CNN weight. We observe that the BSS is nearly the same as the model trained on the CRPS, since the greatest increase in BSS is approximately 0.02. This is much less than the model that uses the indicator weight, where the increase in BSS reached around 0.2 (Figure \ref{fig:bootstrap bss linear models test set}). Presumably, this is due to the fact that the best CNN weight function assigns a lot of weight to moderate wind speeds. Since the training data contains a large number of instances of moderate wind speed, the resulting model closely resembles the one trained on the CRPS.

For the CNN-based model trained using the sharp sigmoid weight in the wCRPS  (Figure~\ref{fig:bootstrap bss transfer weight}, right), we see a different pattern. This model is substantially worse compared to the model trained on the CRPS, except for wind speeds around 15 m/s where it is similar. Differences between sharp sigmoid and best CNN weight functions (Table 4) suggest  that CNNs require information from forecasts of lower wind speeds to post-process extremes. Thus, this information should not be ignored in the weight function to achieve the best results. 

\section{Conclusion and Discussion}\label{sec:conclusion and discussion}

We examined the impact of training linear and non-linear versions of the EMOS model using a variety of wCRPS scores as the loss function instead of the traditional CRPS. We showed that using the wCRPS as a loss function improves performance on high wind speeds in both linear and CNNs models, provided that the weight function is appropriately chosen. For linear models, we observed a strong body-tail trade-off, where improved performance on extremes resulted in reduced performance on the central body of the distribution. To mitigate this problem, we suggested using a class of shifted Gaussian cdfs, to tune how much weight is assigned to extremes compared to the bulk. In combination with the multi-objective hyperparameter optimization algorithm MOTPE, this was shown to be effective in giving the user control on the performance of extremes relative to the performance on the bulk. The body-tail trade-off was not evident for the CNN models. Using the best weight function for CNNs only led to increased performance on extremes without reducing performance on the bulk. This finding is surprising and it would be of interest to investigate whether this phenomenon occurs consistently in other experimental contexts. It is important to note that we performed all the testing of CNNs with bagged versions, since these were able to provide consistent results and were less noisy than the single CNNs.

Another observation is that the optimal weight function does not transfer between models. When the best weight function for linear models was used to train CNNs, performance deteriorated for both the bulk and extremes. Conversely, using the best CNN weight function on linear models resulted in a model that was almost identical to a model trained on a constant weight function, i.e., the CRPS. This indicates that the best choice of weight function in the wCRPS is model-specific.

Additionally, the best choice of parametric distribution was considered for both the linear models and the CNNs. For our data set, the findings indicated that choosing a different distribution did not always lead to substantially better performance, and in tuning there was no consistent choice for a `good' distribution for extremes. This was the case for both the linear models and the CNNs. For both models choosing a different distribution did not affect the performance and in case a mixture distribution was used, the decision on which distribution to use for higher and lower wind speeds seemed more or less random. There could be two reasons why we did not observe any significant difference in the choice of distribution. The first possible reason is the fact that we used a sampling-based approach. This makes the gradients noisier, and estimating the tail behavior of the distribution could have been especially hard since sampling from the tails is unlikely. The second possible reason is that our data set contained ten-minute average wind speeds (rather than ten-minute maximum wind speeds, say), which could benefit the TN. It also could be the case that the choice of distribution is simply not very important in EMOS models, compared to the choice of loss function and the underlying model.

Although using wCRPS can improve performance with the correct choice of weight function, it should first be evaluated whether this is the best approach to take. Tuning the weight function is challenging, since many models have to be trained to find a good set of hyperparameters. For instance, the Brier skill score for a CNN-based model compared to a linear model reference ranged from 0.1 to 0.15, indicating a substantial difference, as shown in Figure \ref{fig:variability cnns}. The increase in Brier skill score for the linear models that solely focus on extremes varied between 0.025 and 0.1, while these models had extremely poor performance on the average wind speeds. For CNNs, the Brier skill score of a model with a well-chosen weight compared to a constant weight function was also approximately 0.05. Therefore it should be carefully considered where computational resources are used, since the change in performance from a simple to a more sophisticated model was much greater than from training on the wCRPS rather than the CRPS. It should be noted that all experiments were done with a CPU, indicating that the computation requirements were not large.

There are several subjects that could be further investigated in future work. First of all, the choice of weight function for CNNs should be studied with a larger search space. In our search space, which contained shifted Gaussian CDFs the optimal weight function resembles an affine function. Therefore, looking at the class of affine functions, or even quadratic function could be interesting. Another promising area of investigation is exploring weighted scoring rules different from the wCRPS, such as the outcome-weighted CRPS \citep{owCRPSIntroduction} and the vertically re-scaled CRPS \citep{Allen2022EvaluatingKernelScores}. For the outcome-weighted CRPS the CRPS is weighted by the outcome and to ensure a proper scoring rule the forecast is evaluated by its weighted representation. The vertically re-scaled CRPS weights the distance between the forecast and the observation, again a different approach from the wCRPS. \citet{CPIT_SamAllen} compare these other weighted variations of the CRPS, as well as their benefits and drawbacks. It would be interesting to see if the best choice of weight function differs between weighted scoring rules. 

\section*{Acknowledgements}

SD was supported by the Dutch Research Council (NWO, Nederlandse Organisatie voor Wetenschappelijk Onderzoek) through the NWO Open Technology grant `Improving probabilistic wind speed forecasts with deep learning'.

\appendix[A]
\appendixtitle{Additional Figures}\label{ap:appendix additional figures}

\begin{figure}[H]
    \centering
    \includegraphics[width=0.9\linewidth]{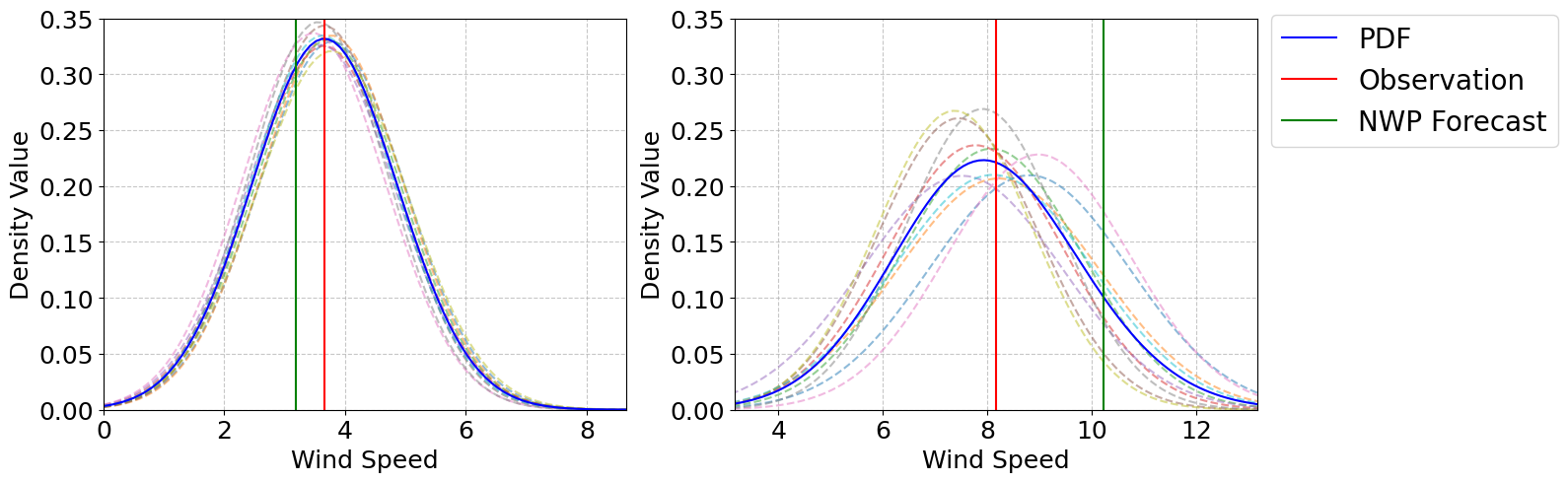}
    \caption{Two forecasts of the CNN combined with bagging (solid curves). The dashed curves show the forecast pdfs of individual CNNs.}
    \label{fig:forecasts pdf bagging 2 figures}
\end{figure}

\begin{table}[H]
\centering
\begin{tabular}{lc}
\toprule
\textbf{Hyperparameter} & \textbf{Range} \\
\midrule
Parametric distribution   & \{Truncated Normal, Log-Normal, Mixture\} \\
$\bm{\mu}$ (weight function)     & [-5, 15]       \\
$\bm{\sigma}$ (weight function)      & [0.0001, 10] (log scale) \\
$\bm{c}$ (weight function) & [0.000001, 1] \\
Optimizer               & \{Adam, SGD\} \\
Learning rate           & [0.0001, 0.03] \\
$\ell_2$-penalty parameter       & [0.00005, 0.1] (log scale) \\
Dense layers       & \{1,2,3,4,5\} \\
Units per dense layer         & [30, 200] (step size 10) \\
Batch size              & \{16, 32, 64, 128, 256, 512, 1024\} \\
\bottomrule
\end{tabular}
\caption{Search space for CNN hyperparameters.}
\label{tab:searchranges cnn hypreopt}
\end{table}

\appendix[B]
\appendixtitle{Station Information}\label{ap:appendix station info}
\begin{center}
\begin{table}[H]
\centering
\resizebox{0.43\textwidth}{!}{\begin{minipage}{1.2\textwidth}
\begin{tabular}{lccc}
\toprule
\textbf{Location} & \textbf{Code} & \textbf{Lon (E°)} & \textbf{Lat (N°)} \\ \midrule
IJMOND                & 209 & 4.518  & 52.464 \\
VOORSCHOTEN           & 215 & 4.436  & 52.140 \\ 
IJMUIDEN              & 225 & 4.555  & 52.462 \\ 
TEXELHORS             & 229 & 4.713  & 52.998 \\ 
DE KOOY               & 235 & 4.781  & 52.927 \\ 
SCHIPHOL              & 240 & 4.790  & 52.317 \\ 
VLIELAND              & 242 & 4.917  & 53.242 \\ 
WIJDENES              & 248 & 5.174  & 52.633 \\ 
BERKHOUT              & 249 & 4.979  & 52.643 \\ 
HOORN (TERSCHLING)    & 251 & 5.346  & 53.391 \\ 
HOUTRIBDIJK           & 258 & 5.401  & 52.648 \\ 
DE BILT               & 260 & 5.180  & 52.099 \\ 
STAVOREN              & 267 & 5.383  & 52.897 \\ 
LELYSTAD              & 269 & 5.521  & 52.458 \\ 
LEEUWARDEN            & 270 & 5.752  & 53.223 \\ 
MARKNESSE             & 273 & 5.888  & 52.702 \\ 
DEELEN                & 275 & 5.872  & 52.055 \\ 
LAUWERSOOG            & 277 & 6.197  & 53.412 \\ 
HEINO                 & 278 & 6.259  & 52.434 \\ 
HOOGEVEEN             & 279 & 6.540  & 52.731 \\ 
EELDE                 & 280 & 6.585  & 53.124 \\ 
HUPSEL                & 283 & 6.657  & 52.068 \\ 
HUIBERTGAT            & 285 & 6.398  & 53.574 \\ 
NIEUW BEERTA          & 286 & 7.149  & 53.194 \\ 
TWENTHE               & 308 & 6.891  & 52.273 \\ 
CADZAND               & 308 & 3.379  & 51.380 \\ 
VLISSINGEN            & 310 & 3.603  & 51.451 \\ 
OOSTERSCHELDE         & 312 & 3.622  & 51.767 \\ 
VLAKTE V.D. RAAN      & 313 & 3.242  & 51.504 \\ 
HANSWEERT             & 315 & 3.998  & 51.446 \\ 
SCHAAR                & 316 & 3.694  & 51.656 \\ 
WESTDORPE             & 319 & 3.841  & 51.225 \\ 
WILHELMINADORP        & 323 & 3.894  & 51.530 \\ 
STAVENISSE            & 324 & 4.006  & 51.568 \\ 
HOEK VAN HOLLAND      & 330 & 4.122  & 51.991 \\ 
THOLEN                & 331 & 4.192  & 51.479 \\ 
WOENSDRECHT           & 340 & 4.342  & 51.448 \\ 
R'DAM-GEULHAVEN       & 343 & 4.433  & 51.892 \\ 
ROTTERDAM             & 344 & 4.435  & 51.925 \\ 
CABAUW                & 348 & 4.936  & 51.969 \\ 
GILZE-RIJEN           & 350 & 4.935  & 51.565 \\ 
HERWIJNEN             & 356 & 5.145  & 51.858 \\ 
EINDHOVEN             & 370 & 5.377  & 51.450 \\ 
VOLKEL                & 375 & 5.707  & 51.659 \\ 
ELL                   & 377 & 5.788  & 51.174 \\ 
MAASTRICHT            & 380 & 5.762  & 50.905 \\ 
ARCEN                 & 391 & 6.196  & 51.497 \\ \bottomrule
\end{tabular}
\end{minipage}}
\caption{Locations of the stations with corresponding codes and coordinates.}
\label{tab:station data}
\end{table}
\end{center}

\appendix[C]

\appendixtitle{Experiments Analytic Expression vs. Sampling-Based Approximation}\label{ap:analytical vs sampling based optimization}

\begin{figure}[h]
    \centering
    \includegraphics[width=0.9\linewidth]{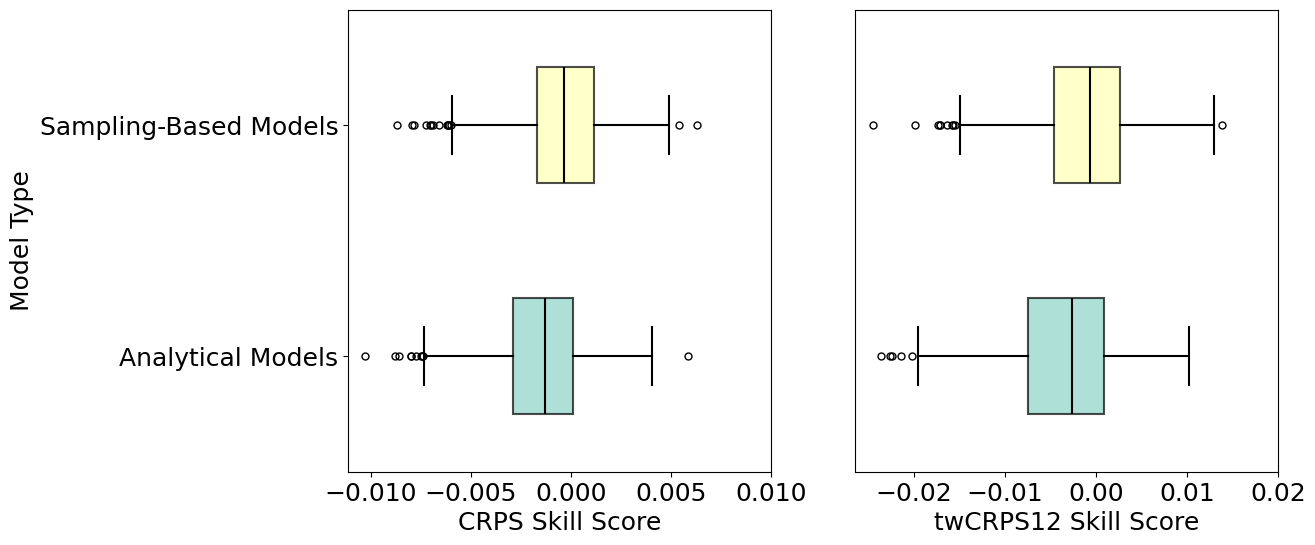}
    \caption{CRPS and twCRPS12 skill scores of the 1000 models trained on the analytical and sampling-based expression of the CRPS.}
    \label{fig:box plot skill scores}
\end{figure}

We performed additional experiments to test the effect of using an analytic expression of the CRPS instead of a sampling-based approximation of the CRPS. These experiments were conducted on a model with the truncated normal distribution, where the CRPS can be computed using the analytic expression 
\begin{align*}
    \text{CRPS}(F,y) = \frac{\sigma}{p^2}\left[sp(2\Phi(s)+p-2) + 2p \phi(s) - \frac{1}{\sqrt{\pi}}\Phi\left(\frac{\mu \sqrt{2}}{\sigma}\right) \right]
\end{align*}
where $p = \Phi(\mu/\sigma)$,$s = (y-\mu)/\sigma$, and $\Phi$ and $\phi$ denote the cdf and pdf of the standard normal distribution \citep{Thorarinsdottir_ProbForecastsWindSpeed_2010}. Except for the loss function, all hyperparameters are fixed, with the same values as in table \ref{tab:linmodelhyperparameters} from section \ref{sec:results}\ref{sec:results linear emos}. We compared the CRPS skill score and twCRPS12 skill score for 1000 models trained on the analytic expression of the CRPS and the sampling-based approximation of the CRPS (see \eqref{eq:wCRPS sampling based estimate} with $v(z)=z$), where the reference model is a model that does not utilize batching and is optimized with the sampling-based approximation of the CRPS. The results can be seen in Figure \ref{fig:box plot skill scores}. To compute the CRPS skill scores for the models in this figure, we used the analytic expression. For the twCRPS12, a sampling-based approximation is used with a sample size of 1000.

We can see that for both the CRPS and the twCRPS12 skill scores, the resulting models are nearly identical. This shows that using the sampling-based approach to compute the CRPS is very reliable, strongly suggesting that using a sampling-based approximation of the wCRPS is a reasonable choice. 


\end{document}